\documentclass[osajnl,preprint,showpacs,superscriptaddress,12pt]{revtex4-1} 

\usepackage{amsmath,amssymb,graphicx}

\begin{document}

\title{An intensity-dependent quantum Rabi model: Spectrum, SUSY partner and optical simulation}

\author{B. M. Rodr\'{\i}guez-Lara}
\affiliation{Instituto Nacional de Astrof\'{i}sica, \'{O}ptica y Electr\'{o}nica \\ Calle Luis Enrique Erro No. 1, Sta. Ma. Tonantzintla, Pue. CP 72840, M\'{e}xico}
\email{bmlara@inaoep.mx}

\begin{abstract}
We study an intensity-dependent quantum Rabi model that can be written in terms of SU(1,1) group elements and is related to the Buck-Sukumar model for the Bargmann parameter $k=1/2$.
The spectrum seems to present avoiding crossings for all valid parameter sets and, thus, may be integrable.
For a degenerate qubit, the model is soluble and we construct an unbroken supersymmetric parter for it.
We discuss the classical simulation of the general model in photonic lattices and show that it presents quasi-periodic reconstruction for a given initial state and parameter set.
\end{abstract}


\maketitle

\section{Introduction} \label{sec:S1}

The Jaynes-Cummings model (JCM) \cite{Jaynes1963p89},
\begin{eqnarray}
\hat{H}_{JC} = \omega \hat{n} + \frac{\omega_{0}}{2} \hat{\sigma_{z}} + g \left( \hat{a} \hat{\sigma}_{+} + \hat{a}^{\dagger} \hat{\sigma}_{-} \right), \label{eq:JaynesCummings}
\end{eqnarray}
is a theoretical model derived from the minimal coupling \cite{Shore1993p1195} between a neutral two-level atom, described by frequency transition $\omega_{0}$ and Pauli matrices $\hat{\sigma}_{j}$, and a quantized cavity field mode, described by the frequency $\omega$ and the creation (annihilation) operators $\hat{a}^{\dagger}$ ($\hat{a}$),  related to the one-atom maser of cavity-quantum-electrodynamics (cavity-QED) \cite{Rempe1987p353,Brune1992p5193}; it can also describe the dynamics of a trapped two-level ion in trapped-ion-QED \cite{Cirac1994p1202} and the coupling of a superconducting qubit interacting with a microwave resonator in circuit-QED \cite{Fink2008p315,De2013p042336}.
The Buck-Sukumar model (BSM) \cite{Buck1981p132}, where the coupling between a two-level system and a quantized field depends on the intensity of the field,
\begin{eqnarray}
\hat{H}_{BS} = \omega \hat{n} + \frac{\omega_{0}}{2} \hat{\sigma_{z}} + g \left( \hat{a} \sqrt{\hat{n}} ~\hat{\sigma}_{+} + \sqrt{\hat{n}} ~\hat{a}^{\dagger} \hat{\sigma}_{-} \right), \label{eq:BuckSukumar}
\end{eqnarray}
is a clever modification of the JCM that leads to a closed form analytic solution.
Its physical realization in the quantum optics laboratory may not be feasible, as it requires a trapped-ion setup driven by a large superposition of field modes \cite{Vogel1995p4214,MoyaCessa2012p229}, but it may be classically simulated in arrays of coupled waveguides \cite{RodriguezLara2013p12888}.
Despite its purely theoretical origin, the BSM \cite{Buck1981p132} and its generalization for qubit ensembles \cite{Sukumar1984p885} have provided analytically tractable models showing periodic decay and revival in the atomic excitation energy \cite{Buck1981p132,Sukumar1984p885}, mean photon number \cite{Singh1982p3206}, and field squeezing parameters \cite{Buzek1989p1151} that has attracted the attention of the quantum optics community.
It is also well known that the field in the BSM can be described by a $su(1,1)$ algebra \cite{Buzek1989p1151,Cordeiro2007p12153,Ng2000p463} and that it is possible to interpolate between the JCM and the BSM by choosing a particular q-deformed algebra for the field \cite{Shanta1992p1301}. 
The inclusion of the so-called counter-rotating terms obviated by the rotating wave approximation into the BSM,
\begin{eqnarray}
\hat{H}_{RBS} = \omega \hat{n} + \frac{\omega_{0}}{2} \hat{\sigma_{z}} + g \left( \hat{a} \sqrt{\hat{n}}  + \sqrt{\hat{n}} ~\hat{a}^{\dagger}  \right) \hat{\sigma}_{x}, \label{eq:BSwithoutRWA}
\end{eqnarray}
reduces the parameter range where the model is well defined to $g < \omega/2$ due to the underlying $su(1,1)$ symmetry \cite{Ng2000p463,Lo1999p557}.

Here we are interested in an intensity-dependent quantum Rabi Hamiltonian that is the simplest generalization of the BSM model without the RWA, 
\begin{eqnarray}
\hat{H} = \omega \hat{n} + \frac{\omega_{0}}{2} \hat{\sigma_{z}} +  g \left( \sqrt{\hat{n}+ 2k} ~ \hat{a}   +  \hat{a}^{\dagger} \sqrt{\hat{n}+2k}  \right) \hat{\sigma}_{x}, \quad k >0, \label{eq:Hamiltonian}
\end{eqnarray}
where a  Bargmann parameter value of $k=1/2$ returns the BSM plus counter rotating terms. 
In quantum electrodynamics this model may be just a theoretical curiosity, but it is of interest in the field of classical optics where it may be possible to classically simulate it with light propagating through a semi-infinite array of coupled waveguides \cite{RodriguezLara2013p12888}.
In the following we will show that this model can be fully written in terms of a su(1,1) algebra due to parity conservation, that it is possible to provide a perturbation theory solution for it in the regime where the qubit transition is negligible, $\omega_{0} \ll g$, that a supersymmetric partner can be given for it in this regime and that both the model and its isospectral partner may be classical simulated by a semi-infinite array of coupled waveguides.

\section{The SU(1,1) model and its spectra} \label{sec:S2}

The Hamiltonian in Eq. (\ref{eq:Hamiltonian}) conserves parity, $\left[ \hat{H}, \hat{\Pi} \right] =0$ with  $\hat{\Pi}= (-1)^{\hat{n}} \hat{\sigma}_{z}$.
This allows us to define two parity subspaces, $\left\{ \vert \pm,j \rangle\right\}$,  such that  $\hat{\Pi} \vert \pm,j \rangle = \pm \vert \pm,j \rangle$ with $ \vert +,j \rangle = \left( \hat{n}^{-1/2}  \hat{a}^{\dagger} \hat{\sigma}_{x} \right)^{j} \vert 0, e \rangle$ and $ \vert -,j \rangle = \left( \hat{n}^{-1/2}  \hat{a}^{\dagger} \hat{\sigma}_{x} \right)^{j} \vert 0, g \rangle$; the states $\vert 0, g \rangle$ and $\vert 0, e \rangle$ correspond to the field in the vacuum state and the qubit in the ground or excited level, in that order.
Thus, Eq. (\ref{eq:Hamiltonian}) becomes the Hamiltonians
\begin{eqnarray}
\hat{H}_{\pm}^{\prime} = \omega \hat{K}_{0} \pm \frac{\omega_{0}}{2} (-1)^{\hat{K}_{0}} + g \left( \hat{K}_{+} + \hat{K}_{-}  \right) - \omega k \pm \frac{\omega_{0}}{2} (-1)^{-k}, \quad k>0, 
\end{eqnarray}
in each parity subspace after defining $\hat{K}_{0} = \hat{n} + k$, $\hat{K}_{+}=\hat{a}^{\dagger} \sqrt{\hat{n}+2k} ~\hat{\sigma}_{x}$ and $\hat{K}_{-}= \sqrt{\hat{n}+2k}  ~\hat{a} ~\hat{\sigma}_{x}$ such that they form the SU(1,1) group, $\left[ \hat{K}_{+}, \hat{K}_{-} \right]= - 2 \hat{K}_{0}$ and $\left[ \hat{K}_{0}, \hat{K}_{\pm} \right]= \pm \hat{K}_{\pm}$ \cite{Barut1980,Novaes2004p351}.
In such a case we can put aside the constant terms and focus on the parity subspace Hamiltonians,
\begin{eqnarray}
\hat{H}_{\pm} = \omega \hat{K}_{0} \pm \frac{\omega_{0}}{2} (-1)^{\hat{K}_{0}} + g \left( \hat{K}_{+} + \hat{K}_{-}  \right).  \label{eq:HSU}
\end{eqnarray}
While in QED it may not make sense, in photonic lattices it is useful to define two regimes where the model is soluble using the qubit frequency as reference: (i) a weak coupling regime where the coupling constant is negligible compared to the qubit frequency, $g \ll \omega_{0}$, where in the case $g / \omega_{0} \rightarrow 0$ the eigenstates of the model are the parity states $\vert \pm, j \rangle$ with energy $E_{\pm,j}= \omega (j+k) \pm \omega_{0} (-1)^{j+k} /2$ and (ii) a deep-strong coupling regime where the qubit frequency is negligible compared to the coupling constant, $g \gg \omega_{0}$, where in the case $\omega_{0} / g \rightarrow 0$ the eigenstates are su(1,1) generalized coherent states,  $\vert \pm, \xi \rangle = S(\xi) \vert \pm , j \rangle$,  with energy $\left( \omega^{2} - 4 g^{2} \right)^{1/2} (j+k)$.
The unitary displacement is given by $S(\xi) = e^{-\xi \left( \hat{K}_{+} - \hat{K}_{-} \right)/2}$ \cite{Gerry1985p2721,Wodkiewicz1985p458} with $\tanh \xi = 2g / \omega$ for our case; note that the displacement parameter, $ \xi = \textrm{arctan} \left(2 g/\omega\right)$, restricts the coupling values for this regime to $g < \omega/2$.
At this point, we can follow an argument identical to that found in \cite{Lo1999p557} and find that despite that the modified evolution operator $\hat{U}_{\pm}= e^{-i \left( \hat{H}_{\pm} + \omega_{0}/2 \right) t}$ is apparently unitary, the value of $\langle \pm, j \vert \hat{U}_{\pm} \vert \pm, k \rangle $ diverges at any finite time for $g \ge  \omega/2$ and, thus, the model seems to be valid just for values of $g<\omega/2$.

\begin{figure}[ht]
\centerline{\includegraphics[scale=1]{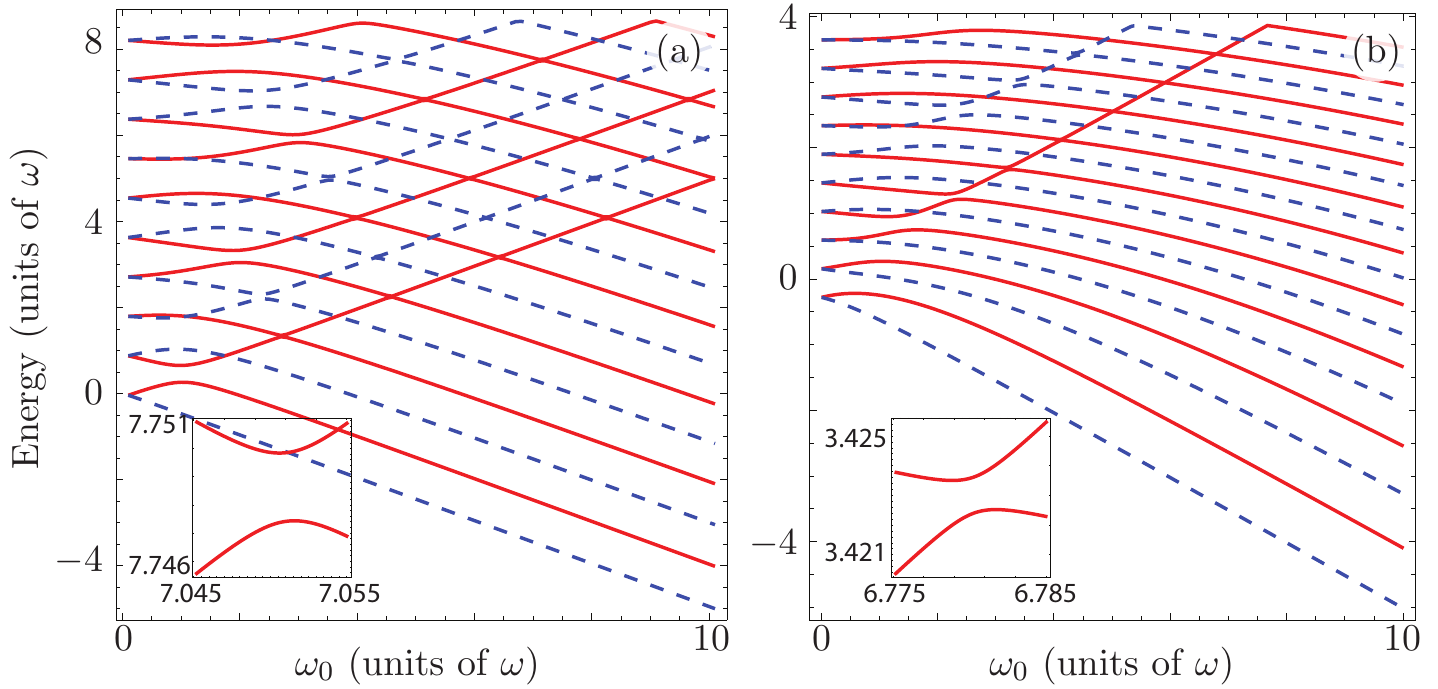}}
\caption{(Color online) The spectra for the positive (red solid lines) and negative (blue dashed lines) parity subspaces of the model $\hat{H}_{\pm}^{\prime}$ with $k=1/2$, that is the BSM plus counter-rotating terms, for variable qubit frequency $\omega_{0}$ with a fixed coupling parameter (a)$g = 0.2 \omega$ and (b)$g = 0.45 \omega$. The insets show typical avoided crossings.} \label{fig:Fig1}
\end{figure}

In any given set of frequencies and coupling parameters, e.g. $\left\{ \omega, \omega_{0}, g\in[0,\omega/2]\right\}$, the model in the parity bases becomes a tridiagonal, real, symmetric semi-infinite matrix which eigenvalues and eigenvectors can be approximated by standard linear algebra methods or discussed analytically following standard methods \cite{Braak2011p100401,Chilingaryan2013p335301}.
Figure \ref{fig:Fig1} shows numerically calculated spectra in the positive and negative parity subspaces for the model Hamiltonian $\hat{H}_{\pm}$ as a function of the qubit frequency. 
The equidistant behavior predicted for the extremes of the weak- and deep-strong-coupling regimes can already be observed.
The spectra shows avoided crossings between the energies of a given parity and crossings between energies of different parity in a similar manner to the spectra of the quantum Rabi Hamiltonian where integrability has been argued on this basis \cite{Braak2011p100401}.

\section{Supersymmetry in the reduced SU(1,1) model}

Let us consider the limiting case of the deep-strong coupling regime where $\omega_{0}= 0$, again this may not make sense while thinking of cavity-, trapped-ion- or circuit-$QED$ but such a model can be produced in photonic lattices \cite{RodriguezLara2013p12888}, and define the unperturbed Hamiltonian 
\begin{eqnarray}
\hat{H}_{0} = \omega \hat{K}_{0} + g \left( \hat{K}_{+} + \hat{K}_{-}  \right).  
\end{eqnarray}
If we define a qubit-field annihilation (creation) operator as $\hat{A} = \alpha \hat{a} + (g/\alpha) \sqrt{\hat{n} + 2k} ~\hat{\sigma}_{x}$ ($\hat{A}^{\dagger} = \alpha \hat{a}^{\dagger} + (g/\alpha) \sqrt{\hat{n} + 2k} ~\hat{\sigma}_{x}$) with parameter $\alpha^{2} = \left( \omega + \sqrt{\omega^{2} - 4 g^{2}} \right)/2$ where the restriction $g < \omega/2$ appears once more, we can write two unbroken supersymmetric partners,
\begin{eqnarray}
\hat{A}^{\dagger} \hat{A} &=& \omega \hat{K}_{0} + g  \left( \hat{K}_{+} + \hat{K}_{-}  \right) - k ~\sqrt{\omega^{2} - 4 g^{2}} ,  \\
\hat{A} \hat{A}^{\dagger} &=& \omega \tilde{K}_{0} + g  \left( \tilde{K}_{+} + \tilde{K}_{-}  \right) + \left( \frac{1}{2}- k \right) \sqrt{\omega^{2} - 4 g^{2}} ,
\end{eqnarray}
where the tilded operators are a different representation of SU(1,1): $\tilde{K}_{0} = \hat{n} + k + 1/2$, $\tilde{K}_{+}= \sqrt{\hat{n}+2k} ~\hat{a}^{\dagger}  ~\hat{\sigma}_{x}$ and $\tilde{K}_{-}= \hat{a} \sqrt{\hat{n}+2k} ~\hat{\sigma}_{x}$. 
Note that both partners are covered by the initial Hamiltonian, Eq. (\ref{eq:Hamiltonian}), for a degenerate qubit because $\tilde{K}_{+}=  ~\hat{a}^{\dagger} \sqrt{\hat{n}+2k + 1}  ~\hat{\sigma}_{x}$ and $\tilde{K}_{-}= \sqrt{\hat{n}+2k + 1} ~ \hat{a} ~\hat{\sigma}_{x}$ .
The two SUSY partners are diagonalized by the displacement $S(\xi)$ defined before and reduce to the following form,  
\begin{eqnarray}
S(-\xi)\hat{A}^{\dagger} \hat{A} S(\xi) &=& \sqrt{\omega^{2} - 4 g^{2}}~ \hat{n} , \quad \Omega_{j} = \sqrt{\omega^{2} - 4 g^{2}}~ j,  \\
S(-\xi)\hat{A} \hat{A}^{\dagger} S(\xi) &=& \sqrt{\omega^{2} - 4 g^{2}}~ \left( \hat{n} +1 \right), \quad \Omega_{j}^{(p)} = \sqrt{\omega^{2} - 4 g^{2}}~ (j+1),
\end{eqnarray}
where it is possible to realize that their spectra are identical, $\Omega_{k} = \Omega_{k-1}^{(p)}$.
A particular case of such unbroken supersymmetric partners has been previously discussed for the parameter set: $k=1/2$, $k=1$, $\omega = 1 - \alpha^{2}$ and $g= -\alpha$ with $\alpha \ne 1$ in the context of photonic isospectral lattices \cite{ZunigaSegundo2014p987}.

\section{Optical simulation}

The optical simulation of quantum Rabi model in arrays of coupled waveguides inscribed by laser damage in fused silica has been proposed and demonstrated experimentally \cite{Crespi2012p163601}.
Nonlinear quantum Rabi models are also feasible of optical simulation \cite{RodriguezLara2013p12888} if care is exerted on the validity of the Hamiltonians and the characteristics of the required lattices \cite{RodriguezLara2014p1784}.
To produce the lattice we follow a standard procedure, which in our case means inserting the general state  $\vert \Psi_{\pm} \rangle = \sum_{j=0}^{\infty}\mathcal{E}_{j}^{(\pm)} \vert \pm ,j \rangle$ and the Hamiltonian $\hat{H}_{\pm}^{\prime}$ into Schr\"odinger equation, to obtain the differential equation sets
\begin{eqnarray}
i \partial_{t} \mathcal{E}_{j}^{(\pm)} = n_{j}^{(\pm)} \mathcal{E}_{j}^{(\pm)} + \gamma_{j-1}  \mathcal{E}_{j-1}^{(\pm)} + \gamma_{j} \mathcal{E}_{j+1}^{(\pm)}, \quad  \mathcal{E}_{-\vert j \vert} =0.
\end{eqnarray}
These sets are equivalent up to a phase with those describing a tight binding photonic lattice where the effective refractive index of the $j$th waveguide is given by $n_{j}^{(\pm)}=\omega j \pm \omega_{0}(-1)^{j}/2$, up to a constant bias refractive index shared by all waveguides, and the coupling between neighbor $j$th and $(j+1)$th waveguides is given by  $\gamma_{j}= g \sqrt{\left( j+1 \right) \left(j + 2k \right)}$, with the Bargmann parameter $k>0$ and the restriction $g<\omega/2$ as discussed before.
The generalities of the optical simulation of quantum phenomena can be found in reviews on the topic \cite{Longhi2009p243,Longhi2011p453,Longhi2013p4884,Rodriguez-Lara2013}.
We want to stress that while the theoretical quantum-optical model requires a semi-infinite array of coupled waveguides, it is possible to cut off the size of the array depending on the initial state to propagate.
This cut off also helps in keeping the photonic lattice experimentally feasible as stronger coupling parameter values require closer waveguides that may prove a complication in the laboratory and produce coupling between second or higher-order neighbors. 

\begin{figure}[ht]
\centerline{\includegraphics[scale=1]{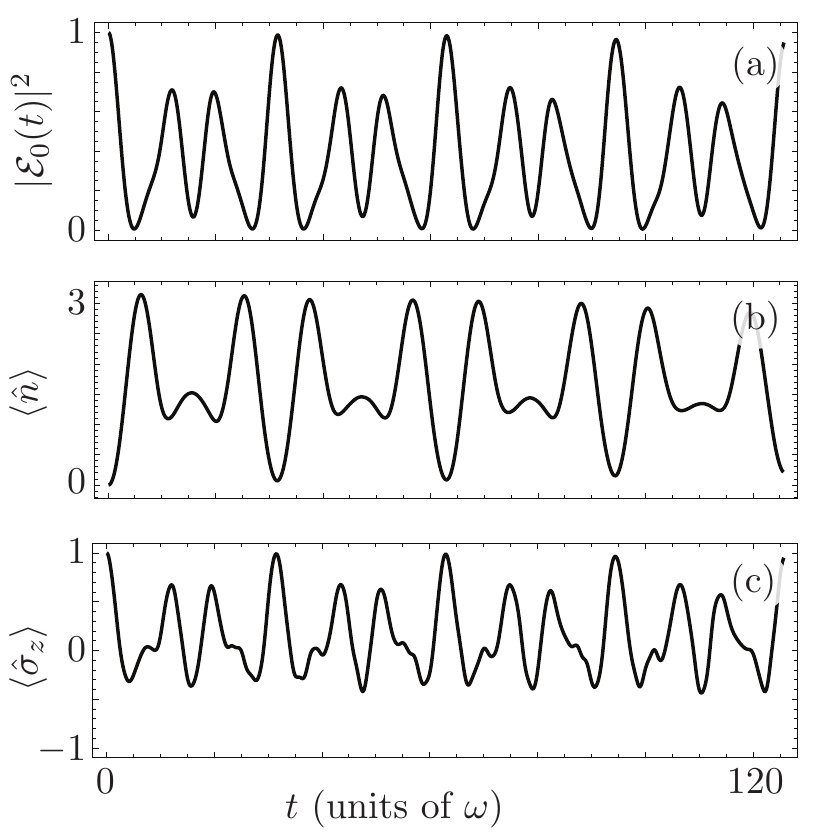}}
\caption {(Color online) Numerical simulation of evolution under $\hat{H}^{\prime}_{+}$ with an initial state $\vert 0,e\rangle$ and parameter set $\{ \omega, \omega_{0}, g, k \} = \{ \omega, 3 \omega / 4, 4\omega/10, 1/2 \}$. (a)Intensity at the zeroth waveguide, (b) mean photon number equivalent to the intensity barycenter, (c) mean atomic excitation energy equivalent to the total intensity in odd waveguides minus the total intensity in even waveguides. }  \label{fig:Fig2}
\end{figure}

Figure \ref{fig:Fig2} shows the time evolution of the initial state $\vert \psi(0) \rangle = \vert 0,e\rangle$ under the dynamics imposed by $\hat{H}^{\prime}_{+}$ as a classical simulation provided by light impinging the zeroth waveguide of a photonic lattice with the parameter set $\{ \omega, \omega_{0}, g, k \} = \{ \omega, 3 \omega / 4, 2\omega/5, 1/2 \}$ and a lattice size of two hundred waveguides.
Quasi-periodical  $\sim 10 \pi$ returns to a state close to the initial state can be observed in the intensity of the zeroth waveguide, $\vert \mathcal{E}_{0}(t) \vert^{2}$, mean photon number, $\langle \hat{n}(t) \rangle = \sum_{j} j \vert \mathcal{E}^{(+)}_{j}(t) \vert^{2}  $ which is equivalent to the barycenter of the intensity, and  mean atomic excitation energy, $\langle \hat{\sigma}_{z}(t) \rangle = \sum_{j} \left[ \vert \mathcal{E}^{(+)}_{2j + 1}(t) \vert^{2} -  \vert \mathcal{E}^{(+)}_{2j}(t) \vert^{2} \right]$ which is equivalent to the difference between the total intensity at odd and even waveguides. 
This is an interesting phenomenon that we were not expecting in the model for such a high coupling parameter and should be probed in the future.

\section{Conclusions} \label{sec:S6}

We have proposed an intensity-dependent quantum Rabi model with an underlying parity and SU(1,1) symmetry.
In the case $k=1/2$ our model reduces to the Buck-Sukumar model plus counter-rotating terms.
As expected from the behaviour of the Buck-Sukumar model, our model seems to be invalid for coupling factors of $g \ge 2$.
The behavior of the spectra is similar to the quantum Rabi model; that is, avoided crossings in spectral branches belonging to the same parity and crossings between spectral branches belonging to different parities.  
In the special case of degenerate qubit frequency, $\omega_{0} = 0$, it is straightforward to diagonalize the model with generalized SU(1,1) coherent states.
It is also possible to provide qubit-field creation and annihilation operators that fulfill the commutator for the field and allows us to construct an unbroken SUSY partner for it; the SUSY partners correspond to Bargmann parameters $k$ and $k_{p}= k+1/2$.
This gives a recipe to a class of isospectral photonic lattices.
Finally, we discussed the classical simulation of the full theoretical model in finite arrays of coupled photonic waveguides and showed by numerical simulation that it is possible to find quasi-periodic reconstruction for a given initial state in the full intensity-dependent quantum Rabi model for a given parameter set.



\bibliographystyle{osajnl}

\begin{thebibliography}{10}
\newcommand{\enquote}[1]{``#1''}

\bibitem{Jaynes1963p89}
E.~T. Jaynes and F.~W. Cummings, Proc.
  IEEE \textbf{51}, 89 -- 109 (1963).

\bibitem{Shore1993p1195}
B.~W. Shore and P.~L. Knight, J. Mod.
  Optics \textbf{40}, 1195--1238 (1993).

\bibitem{Rempe1987p353}
G.~Rempe, H.~Walther, and N.~Klein, Phys. Rev. Lett. \textbf{58}, 353 -- 356
  (1987).

\bibitem{Brune1992p5193}
M.~Brune, S.~Haroche, J.~M. Raimond, L.~Davidovich, and N.~Zagury,
   Phys. Rev. A \textbf{45}, 5193 -- 5214 (1992).

\bibitem{Cirac1994p1202}
J.~I. Cirac, R.~Blatt, A.~S. Parkins, and P.~Zoller,  Phys. Rev. A \textbf{49},
  1202 -- 1207 (1994).

\bibitem{Fink2008p315}
J.~M. Fink, M.~G\"oppl, M.~Baur, R.~Bianchetti, P.~J. Leek, A.~Blais, and
  A.~Wallraff, Nature \textbf{454}, 315 --318
  (2008).

\bibitem{De2013p042336}
A.~De and R.~Joynt,  Phys. Rev. A \textbf{87}, 042336 (2013).

\bibitem{Buck1981p132}
B.~Buck and C.~V. Sukumar,  Phys. Lett. \textbf{81}, 132 --
  135 (1981).

\bibitem{Vogel1995p4214}
W.~Vogel and R.~L. de~Matos~Filho, Phys. Rev. A \textbf{52}, 4214 -- 4217 (1995).

\bibitem{MoyaCessa2012p229}
H.~Moya-Cessa, F.~Soto-Eguibar, J.~M. Vargas-Mart\'{\i}nez, R.~Ju\'arez-Amaro,
  and A.~Z{\'u}{\~n}iga-Segundo, Phys. Rep. \textbf{513}, 229 -- 261 (2012).

\bibitem{RodriguezLara2013p12888}
B.~M. Rodr{\'\i}guez-Lara, F.~Soto-Eguibar, A.~Z. C\'ardenas, and H.~M.
  Moya-Cessa,  Opt. Express \textbf{21}, 12888 -- 128981
  (2013).

\bibitem{Sukumar1984p885}
C.~V. Sukumar and B.~Buck,  J. Phys. A: Math. Gen. \textbf{17}, 885 -- 894 (1984).

\bibitem{Singh1982p3206}
S.~Singh,  Phys. Rev. A \textbf{25}, 3206 -- 3216 (1982).

\bibitem{Buzek1989p1151}
V.~Buzek,  J. Mod. Opt. \textbf{36}, 1151 -- 1162 (1989).

\bibitem{Cordeiro2007p12153}
F.~Cordeiro, C.~Providencia, J.~da~Providencia, and S.~Nishiyama, J. Phys. A Math. Theor. \textbf{40}, 12153--12160 (2007).

\bibitem{Ng2000p463}
K.~M. Ng, C.~F. Lo, and K.~L. Liu,
  Physica A \textbf{275}, 463 -- 474 (2000).

\bibitem{Shanta1992p1301}
P.~Shanta, S.~Chaturdervi, and V.~Srinivasan,
  J. Mod. Opt. \textbf{6}, 1301 -- 1308 (1992).

\bibitem{Lo1999p557}
C.~F. Lo, K.~L. Liu, and K.~M. Ng, 
  Physica A \textbf{265}, 557 -- 564 (1999).

\bibitem{Barut1980}
A.~O. Barut and R.~Raczka, \emph{Theory of group representations and
  applications} (PWN Polish Scientific Publishers, 1980).

\bibitem{Novaes2004p351}
M.~Novaes, Rev. Bras. Ensino Fis.
  \textbf{26}, 351 -- 357 (2004).

\bibitem{Gerry1985p2721}
C.~C. Gerry, Phys. Rev. A
  \textbf{31}, 2721 -- 2723 (1985).

\bibitem{Wodkiewicz1985p458}
K.~W\'odkiewicz and J.~H. Eberly, J. Opt. Soc. Am. B \textbf{2}, 458 -- 466 (1985).

\bibitem{Braak2011p100401}
D.~Braak, Phys. Rev. Lett.
  \textbf{107}, 100401 (2011).

\bibitem{Chilingaryan2013p335301}
S.~A. Chilingaryan and B.~M. Rodr\'iguez-Lara,  J. Phys A: Math. Theor. \textbf{46}, 335301 (2013).

\bibitem{ZunigaSegundo2014p987}
A.~Z\'uniga-Segundo, B.~M. Rodr\'iguez-Lara, D.~J.~F. C., and H.~M. Moya-Cessa,
   Opt. Express
  \textbf{22}, 987 -- 994 (2014).

\bibitem{Crespi2012p163601}
A.~Crespi, S.~Longhi, and R.~Osellame,  Phys. Rev. Lett. \textbf{108}, 163601 (2012).

\bibitem{RodriguezLara2014p1784}
B.~M. Rodr{\'\i}guez-Lara, F.~Soto-Eguibar, A.~Z. C\'ardenas, and H.~M.
  Moya-Cessa,  Opt. Express
  \textbf{22}, 1784 -- 1787 (2014).

\bibitem{Longhi2009p243}
S.~Longhi,  Laser
  Photon. Rev. \textbf{3}, 243 -- 261 (2009).

\bibitem{Longhi2011p453}
S.~Longhi,  Appl. Phys. B \textbf{104}, 453 -- 468 (2011).

\bibitem{Longhi2013p4884}
S.~Longhi,  Opt. Lett. \textbf{38}, 4884 -- 4887 (2013).

\bibitem{Rodriguez-Lara2013}
B.~M. Rodr\'iguez-Lara, F.~Soto-Eguibar, and D.~N. Christodoulides,  arXiv: 1311.3694
  [physics.optics] (2013).

\end{thebibliography}

\end{document}